\documentclass{article}
\usepackage[T1]{fontenc} 
\usepackage{fouriernc}
\usepackage{microtype} 
\usepackage{graphicx}
\usepackage{float}
\usepackage[caption = true]{subfig}
\usepackage[superscript]{cite}

\usepackage[hmarginratio=1:1,top=32mm,columnsep=20pt]{geometry} 
\usepackage{multicol} 
\usepackage{booktabs} 
\usepackage{float} 
\usepackage{hyperref} 

\usepackage{paralist} 

\usepackage{abstract} 

\usepackage{titlesec} 
\renewcommand\thesection{\Roman{section}} 
\renewcommand\thesubsection{\Roman{subsection}} 
\titleformat{\section}[block]{\large\scshape\centering}{\thesection.}{1em}{} 
\titleformat{\subsection}[block]{\large}{\thesubsection.}{1em}{} 

\usepackage{fancyhdr} 
\title{\vspace{-15mm}\fontsize{24pt}{10pt}\selectfont\textbf{Ecological metrics of diversity in understanding social media}} 

\author{
\large
\textsc{Chris von Csefalvay FRSA}\thanks{MA (Oxon), BCL (Oxon), Fellow of the Royal Society of Arts. Technical architect, RB plc, 215 Bath Rd, Slough SL1 4AA. E-mail: \href{mailto:chris@chrisvoncsefalvay.com}{chris@chrisvoncsefalvay.com}. All opinions contained in this paper are the author's own and do not necessarily reflect the positions of any institution the author is affiliated with.}\\[2mm] 
\vspace{-5mm}
}
\date{}

\begin{document}

\maketitle 

\thispagestyle{fancy} 


\begin{abstract}

\noindent Topical discussion networks (TDNs) are networks centered around a discourse concerning a particular concept, whether in real life or online. This paper analogises the population of such networks to populations encountered in mathematical ecology, and seeks to evaluate whether three metrics of diversity used in ecology - Shannon's $H'$, Simpson's $\lambda$ and $E_{var}$ proposed by Smith and Wilson give valuable information about the composition and diversity of TDNs. It concludes that each metric has its particular use, and the choice of metric is best understood in the context of the particular research question. 

\end{abstract}


\begin{multicols}{2} 

\section{Introduction}

In analysing a a topical discussion network (TDN),\cite{bruns2011new,gonzalez2010structure,highfield2011challenges} such as tweets mentioning the same hashtag or same keyword, an important question for the understanding of information flows is the source diversity among various contributors to the conversation. In other words, what are the 'market shares' attributable of individual sources? Is the conversation dominated by a small number of contributors or is it a balanced exchange with a large number of actors holding conversations at relative parity? 

This paper examines the utility of three metrics - Shannon's $H'$, the Simpson $\lambda$ metric and the $E_{var}$ metric proposed by Smith and Wilson (1996) - in understanding TDNs in social media. In particular, it is attempting to ascertain what each of these metrics reveal about information flows within a TDN, and to what extent these metrics can be adapted to TDN analysis. In this sense, a TDN is regarded as an analogue of an entire ecosystem, with each contributor being a distinct 'species'. Their contributions - whether Facebook posts, tweets, Pinterest pins or Instagrams - are equivalent to a species's weight in an ecosystem. Analogous expansion of metrics in mathematical ecology is by no means unknown. Indeed, the same expansion was carried out, independently, by Herfindahl and Hirschman, in creating the mathematically identical Herfindahl-Hirschman index used in antitrust law as a proxy of market concentration.\cite{rhoades1993herfindahl,calkins1983new,miller1982herfindahl,cohen1983herfindahl,hannan1997market} This paper asks, principally, whether such an expansion would yield the same new insights that it yielded in economics. In particular, it asks whether the metrics used in mathematical ecology translate to the rather different field of analysing networks in human interaction, where the number of 'species' can often be quite large.

This paper considers a particular manifestation of TDNs, namely Twitter hashtags. This is partly owing to the ethical issues involved in using a \emph{prima facie} closed social network, such as Facebook, in the research of often contentious political issues, as well as due to the relative ease by which the data necessary for such research is available from Twitter's API. While the results do not indicate any \emph{prima facie} source dependence, further research using other information sources is clearly warranted. Insights from such studies may lead to a better understanding of how political discourse is acted out in the online public space and how such discourse can be described in statistical terms.


\section{Background}
\label{sec:background} 

\subsection{Defining metrics} 
\label{sub:defining_metrics}


Consider a TDN, denoted as $\Theta$, with $R$ contributors $C_1 \cdots C_R$ and a set ${T}$ of $V$ elements comprising all contributions. Let the number of contributions by contributor $C_n$ be denoted as $t_n$. We can then construct a probability space in which

\begin{enumerate}
	\item the sample space $\Omega$ comprises all contributions, i.e. 
		\begin{equation}
			\Omega = \left\{\omega | \omega \in T\right\}
		\end{equation}
    \item the $\sigma$-algebra $\mathcal{F}$ the set of all subsets in the sample space, i.e. $2^\Omega$, and
	\item the probability measure that a contribution selected at random will be from a contributor $C_j$ as
		\begin{equation}
			\mathcal{P}(\omega_j) = \frac{t_j}{\sum\limits_{i = 1}^{R} t_i}.
		\end{equation}
\end{enumerate}

In the following, three metrics will be defined for any given TDN $\Theta$ consisting of $R$ distinct contributors and $N$ distinct contributions where the number of contributions by contributor $j$ is denoted as $t_j$ and the proportion of the contributions by that user is denoted as $p_j$.

\subsection{Shannon-Weaver index} 
\label{sub:shannon_weaver_index}

Measuring the diversity of multispecies populations using the concept of 'information content' gained traction in the late 1950s, following the work of Claude Shannon on the information entropy of communication.\cite{shannon1993collected} Under the 'information content' understanding of species diversity, the diversity of a multispecies population is equivalent to the uncertainty of finding an individual of species $i$ when randomly selecting an individual from the population $P$.\cite{pielou1966shannon,magurran1988ecological,pielou1966measurement} In other words, the 'information content' of an individual within the population depends on the population's information (or species-occurrence) entropy.

Pielou (1966) distiguishes two definitions of information content that were initially prevalent - one version, which he credits to Brillouin (1960), is usable where the total number of individuals, i.e. the size of $V$ in our case, as well as the richness $R$ of the population, is known. In that case, the diversity of a population can be expressed as 

\begin{equation}
	H = \frac{1}{N} \ln N\! - \sum_{j = 1}^{R} \ln t_j\!
\end{equation}

Brillouin's formula, however, does not avail us in the all too frequent case where the size of the entire population is unknown or unknowable. In that scenario, the true population diversity cannot be calculated. It can, however, be estimated from a sample. For a sample of richness $R$, the estimated population diversity $H'$ is given by

\begin{equation}
	H' = - \sum_{j = 1}^{R} p_j \ln p_j
\end{equation}

where $p_i$ denotes the proportion of species $j$ in the sample. From (2) follows that for a contributor $j$, the proportion of his tweets can be represented as

\begin{equation}
	\frac{t_j}{\sum\limits_{i = 1}^{R} t_i} = \frac{t_j}{N}
\end{equation} 

Consequently, $H'$ can be calculated as

\begin{equation}
	H' = - \sum_{j = 1}^{R} \frac{t_j}{N} \ln \frac{t_j}{N}
\end{equation}

Shannon's index is one of diversity, i.e. it is strongly richness-dependent and it measures primarily the entropic dissimilarity of the population rather than the even distribution. A derivative of the index, known sometimes as Shannon's evenness metric $J'$ exists, which is defined as

\begin{equation}
	J' = \frac{H'}{\ln R} = \frac{- \sum\limits_{j = 1}^{R} \frac{t_j}{N} \ln \frac{t_j}{N}}{R}
\end{equation}


\subsection{The Simpson lambda index} 
\label{sub:the_inverse_simpson_index}

The Simpson index, usually denoted by $\lambda$, is the probability that two entities drawn at random from the population, with replacement, will be of the same species.\cite{hunter1988numerical} For the probability space associated with the TDN $\Theta$ as described above, it can be calculated as

\begin{equation}
	\lambda = \sum_{j = 1}^{R} p_j^2 = \sum_{j = 1}^{R} \Big(\frac{t_j}{N}\Big)^{2}
\end{equation}

In economics, this index is known as the Herfindahl-Hirschman Index (HHI), and serves to quantify market concentration. Since its inception in 1950, a year after Simpson's independent discovery of the same concept in mathematical ecology, the Herfindahl-Hirschman Index has become the gold standard proxy for market power, and forms the basis of the US Department of Justice's analysis of possible anticompetitive effects of mergers. The index "accounts for the number of firms in the market, as well as the concentration, by inforporating the relative size (that is, market share) of all firms in the market".\cite{rhoades1993herfindahl} Its transferability indicates that the phenomenon it measures, namely relative concentration, is not specific to the domain of its origin, but rather describes accumulative relationships and diversity in various domains.


\subsection{The $E_{var}$ metric} 
\label{sub:the_e_var_metric}

It has been observed that the number of distinct species, known as richness (represented as $R$ in the above model of the TDN $\Theta$), has a significant impact on diversity metrics.\cite{alatalo1981problems} The richness problem scales as we transfer the metrics from population ecology to social interactions. In many of the samples that will be discussed in our research, $R$ will be in the thousands - indeed, the largest sample set that will be considered has a richness of almost 150,000 distinct contributing entities. Furthermore, the previous indices primarily focused on diversity, whereas that only delivers part of the picture. Evenness, too, plays a significant role in understanding a TDN. To take account of evenness, multiple approaches have been proposed. One was to adapt the index proposed by McIntosh (1967) for the measurement of species diversity\cite{mcintosh1967index}, into an index of evenness. The index proposed by Pielou would, for the TDN $\Theta$ as described above, be calculated as

\begin{equation}
	E_{McI}(\Theta) = \frac{N - \sqrt{\sum\limits_{i = 1}^{R} t_i^2}}{N - \frac{N}{\sqrt{R}}}
\end{equation}

Smith and Wilson (1996) propose a new metric, $E_{var}$, which is intuitively based on the variance of the logarithm of each species's population.\cite{smith1996consumer} It uses a trigonometric transformation first used by Alatalo (1981) to reduce the result to a value in radians.\cite{alatalo1981problems} 
\small
\begin{equation}
	E_{var}=1-\frac{2}{\pi}\arctan\Bigg\{\frac{1}{R} \sum\limits_{i = 1}^{R} \Bigg(\ln(p_i)-\sum\limits_{j=1}^R \frac{\ln(p_j)}{R}\Bigg)^2\Bigg\}
\end{equation}
\normalsize
Smith and Wilson have proved this metric to be independent of species richness for all values of $R$, as well as its sensitivity to changing the abundance of the most minor species. The $E_{var}$ metric will be considered alongside the Shannon-Weaver and Simpson complement indices as a non-$R$ sensitive metric.


\subsection{Adaptation of the indices: divergences and challenges} 
\label{sub:adaptation_of_the_indices_divergences_and_challenges}

A primary feature of these indices is that they were developed with classical population ecology in mind.\cite{magurran2004measuring,mason2003index} Where they have been adapted, they have generally applied only to a small number of entities - thus, for instance, in the context of the Herfindahl-Hirschman index, the typical number of undertakings to consider does not exceed a few dozen.\cite{miller1982herfindahl} As such, the adaptation to the economic context was reasonably unproblematic, as given narrow industry definitions, most of the contribution to the Herfindahl-Hirschman index was by a relatively small number of undertakings, approximately on par with (or even smaller than) the number of distinct species in an ecological study. Even where that was not the case, a capping mechanism was implemented, calculating the index for, conventionally, the top 20 or top 50 undertakings.\cite{bailey1971optimal,weinstock1982using,marfels1975bird}

The issue of much higher richness - almost 150,000 in our largest sample, from the \texttt{\#gamergate} hashtag - means richness-sensitive metrics, such as Simpson's $\lambda$ and $H'$, are more at risk of not reflecting real diversity than the same metrics would be in the relatively confined domain of ecology or competition economics, where the upper bound of richness was much smaller. To alleviate any issues with adaptation, the present research employs two methodologies. First, a richness-insensitive metric $E_{var}$ is employed in addition to the richness-sensitive metrics.\cite{smith1996consumer} Second, the metrics were calculated, alongside the entire sample, for the top quintile and the top decile of contributors, by proportion of contributions.  



\section{Methodology} 
\label{sec:methodology}

For the purpose of this research, a sample of twelve TDNs centered around a range of topics were examined. Using a high-throughput automated retrieval engine written in Python that interacts with Twitter using a RESTful API, samples of various sizes were obtained from a number of hashtags selected for topical unambiguity (a single hashtag or search term should retrieve most, if not all, of a community but not retrieve results concerned with a different meaning of the word), as well as to cover each typological class of 'virtual community' proposed by Porter.\cite{porter2004typology} The samples were collected throughout the period from October to December 2014.

The 12 hashtags ultimately selected comprise a range of subject areas, popularity and sample sizes. The samples were stored in a \texttt{MongoDB} instance and queried by a custom \texttt{Python} script, which then calculated the pertinent metrics using \texttt{pandas} as a data abstraction. The metrics were calculated by reference to individual Twitter users' identifiers, rather than their user names or 'screen names', which are both mutable whereas user identifiers are assigned at the time of account creation and are globally unique both across Twitter users and across time.

\end{multicols}
\begin{table}[H]
\caption{Hashtags included in the research}
\centering
\begin{tabular}{llcccc}
\toprule
\multicolumn{2}{c}{Hashtag}\\
\cmidrule(r){1-2}
Name & Subject type & $N$ & $R$ & $\overline{p_i}\times10^6$ & $\sigma{p_{i}}\times10^6$ \\
\midrule

\texttt{\#auspol} & Politics & $206,040$ & $25,410$ & $39.3544$ & $18.0504$ \\
\texttt{\#blacklivesmatter} & Politics & $216,097$ & $101,539$ & $9.8484$ & $36.3133$ \\
\texttt{\#cashinin} & Politics & $3,682$ & $1,258$ & $794.9212$ & $2103.0147$ \\
\texttt{\#dataviz} & Professional & $5,079$ & $3,236$ & $309.0175$ & $641.6421$ \\
\texttt{\#ferguson} & Politics & $354,548$ & $128,800$ & $7.7648$ & $41.1617$ \\
\texttt{\#gamergate} & Entertainment & $3,711,580$ & $146,472$ & $6.8273$ & $247.8704$ \\
\texttt{\#mtvstars} & Entertainment & $103,400$ & $26,638$ & $37.5406$ & $76.3736$ \\
\texttt{\#p2} & Politics & $88,594$ & $24,085$ & $41.5660$ & $23.4973$ \\
\texttt{\#rstats} & Professional & $1,071$ & $645$ & $1,550.4202$ & $2.3789$ \\
\texttt{\#startup} & Professional & $36,100$ & $7,550$ & $132.4515$ & $36.0704$ \\
\texttt{\#tcot} & Politics & $567,763$ & $85,717$ & $11.6663$ & $66.5921$ \\
\texttt{\#uniteblue} & Politics & $59,280$ & $15,496$ & $64.5327$ & $14.4400$ \\

\bottomrule
\end{tabular}
\end{table}
\begin{multicols}{2}

After determining the sample size ($N$), richness ($R$) and average $p_i$ for each hashtag sample, the three metrics that form the basis of this study have been calculated for each of the hashtags using a Python implementation of the calculations. The metrics were then separately calculated for two subpopulations, namely the top quintile and top decile of users, respectively, to assist with the estimation of each metric's sensitivity to $R$.

\end{multicols}
\begin{figure}
	\subfloat[Shannon's $H'$]{\includegraphics[width = 2in]{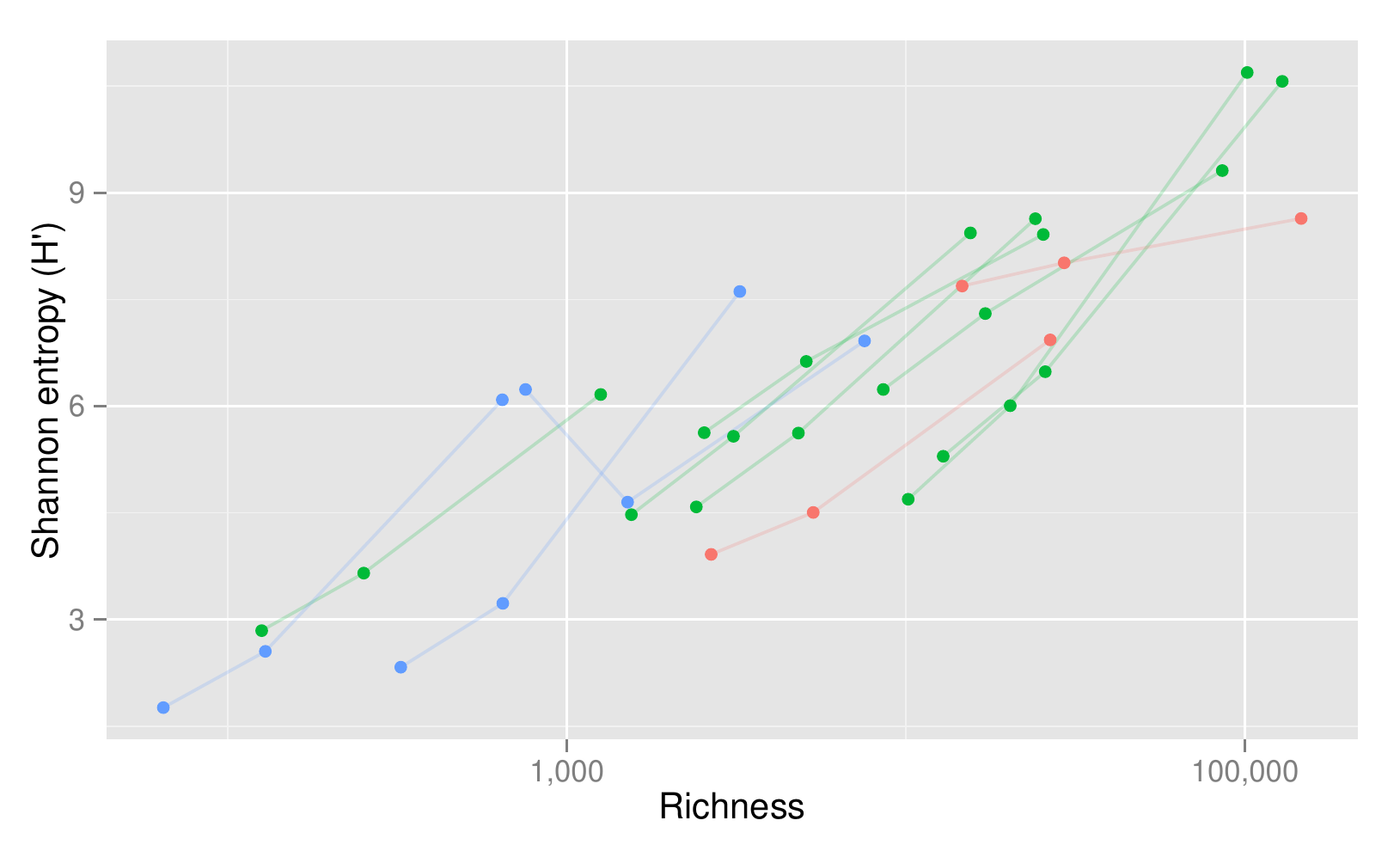}} 
	\subfloat[Simpson's $\lambda$]{\includegraphics[width = 2in]{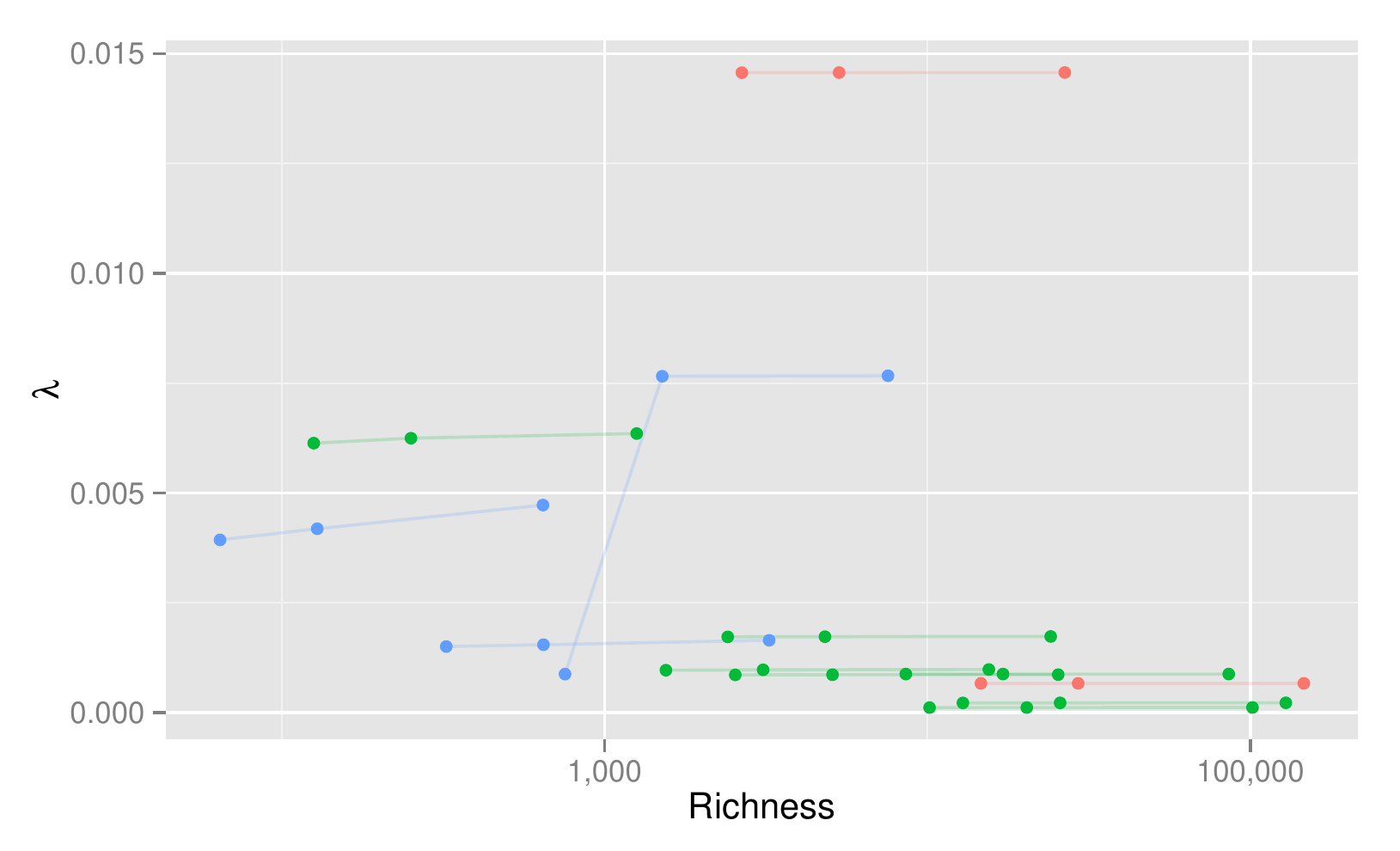}}
	\subfloat[$E_{var}$]{\includegraphics[width = 2in]{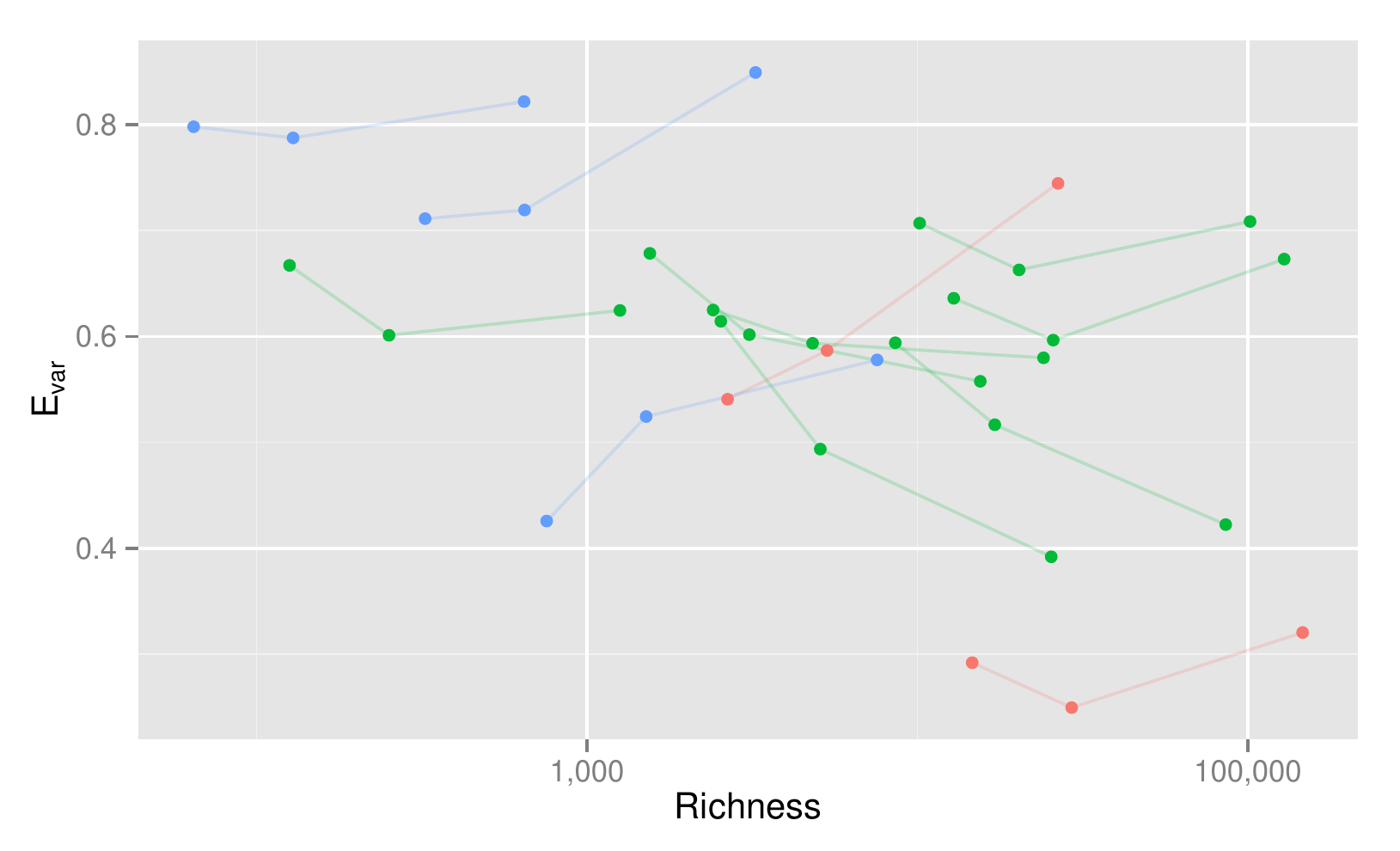}}
	\caption{Metrics calculated for the sampled TDNs. Subsamples derived from the same hashtag (total, first quintile and first decile) are connected by lines. The colours identify the TDN's classification (red = entertainment, green = politics, blue = professional).}
	\label{metrics-comparison}
\end{figure}
\begin{multicols}{2}

In addition, to test each metric's dependence on $R$, the Pearson product-moment correlation was computed for each of three subsamples of each hashtag's population, namely the whole hashtag, the top quintile and the top decile. The examination of the correlations found that $H'$ strongly positively correlates with $R$ ($r = 0.692$, 95\% CI: $0.471$ - $0.832$) and weakly negatively correlates to each of $\lambda$ ($r = -0.253$, 95\% CI: $-0.553$ - $0.059$) and $E_{var}$ ($r = -0.275$, 95\% CI: $-0.553$ - $0.059$). Of these correlations, only that between $R$ and $H'$ is statistically significant at $p < 0.05$. This indicates that at the sample sizes considered, both $\lambda$ and $E_{var}$ deliver accurate results that are not statistically significantly influenced by the richness of the sample. Figure~\ref{metrics-comparison} shows the relationship between the selected metrics and the richness of the sample.

In agreement with the research by Smith and Wilson (1996), this examination of the evenness and diversity metrics concludes that $H'$ is significantly affected by the richness $R$ in the sample.\cite{smith1996consumer} Following the distinction drawn by Pielou (1977), a metric of diversity ought ideally to measure only one of the constituent components of diversity, namely either richness or evenness, but not of both. Thus, a metric significantly affected by richness would be an unsuitable metric of evenness, and vice versa.\cite{pielou1977mathematical} Based on this, a methodological role for each of the metrics emerges:

\begin{enumerate}
	\item \textbf{Shannon's $H'$} is unsuitable for comparisons between TDNs of significantly divergent $R$, just as they would not be suitable for comparing biomes with significantly divergent numbers of species. This is not necessarily a drawback, however, where the comparison requires richness to be taken into account. To the extent that one is concerned with the relative likelihood of encountering divergent opinions, the richness of a sample is hardly irrelevant. Indeed, the richness of a TDN might become relevant when considering such questions as whether the conversation is subject to the dominance of a few or a widely participatory marketplace, even if it does not allow comparison of evenness independently of richness. It would, therefore, be premature to discard it as useless - within the realm of social media analysis, $H'$ could serve to distinguish low-participation, niche TDNs from TDNs where participation is wide and relatively even, with the caveat that it will be less sensitive to unevenness as richness increases.
	\item With the exception of a single outlier (the \texttt{\#startup} hashtag), \textbf{Simpson's $\lambda$} appears to be the least affected by divergences in richness, both in the case of subsamples drawn from the same sample and in the case of inter-sample differences. This indicates that $\lambda$ is a good metric for measuring and comparing evenness and measuring dominance in a way that is less sensitive to $R$ and (quadratically) more sensitive to the existence of entities with a larger share of the conversation - in this instance, it excelled at identifying the outlier, the \texttt{\#mtvstars} hashtag, which was marked by the strong participation of a few significant sources (mainly media outlets and celebrity bloggers), a distinction other metrics did not pick up on. As such, where the research question seeks to identify relative imbalances between the largest few and the remainder of the sample and thereby pinpoint situations of unusual dominance, the quadratic amplification of such dominance by the $\lambda$ metric is a helpful mathematical tool.
	\item The \textbf{$E_{var}$ metric} is also relatively unaffected by changes in $R$, although not to the extent that Simpson's $\lambda$ is in most cases. It is inferior to $\lambda$ in discerning dominance by a small number of highly dominant entities. It does, however, deliver superior performance in discerning the relative evenness between each contributor's share in the conversation in a way that is largely sensitive to changes in the share of the conversation held by the lowest-contributing contributors. As such, as each subsample is expanded, the expansion yields either an increase or a decrease in $E_{var}$, reflecting the influence of the less dominant contributors' shares on the index, whereas such expansion does not affect $\lambda$ as most of a sample's $\lambda$ is determined by its first decile (indeed, often only a fraction thereof).

\end{enumerate}


\section{Results} 
\label{sec:results}

What do metrics of concentration and evenness teach us about a TDN and the people who contribute to it? As the discussion in the previous chapter has shown, each of the three metrics considered in this paper has a particular role in discerning the diversity and evenness of a sample derived from a TDN. In mathematical ecology, the concept of diversity, dominance and evenness serves to understand the interrelationship between various species of various abundances each. Do a few species dominate or do a large number of different species share the resources available to the population? Do individuals fall into species with relatively even probabilities or is there a distinct 'fall-off'? In this sense, the relative dominance relationships between individual species within a population can be categorised and understood based purely on mathematical indicators. In the context of a social network, such as a TDN, the issue is slightly different. For one, the constraining factor is slightly different. In a TDN, voices compete for a share of the conversation. Each contribution comes at a cost in terms of time, energy and various system-provided maxima of daily or hourly contribution rates (e.g. Twitter's maximum of 2,400 messages). As such, the contributions represent not merely how much an individual contributor's opinion adds to the whole of the conversation but also a measure of his or her expenditure in terms of time and the relatively limited resource of daily tweets to the conversation. In this sense, higher $p_i$ translates not only to a louder, more influential voice but also to a more significant expenditure of a relatively scarce resource on participating in a TDN. Consequently, a TDN with high concentration indicates the presence of agents who are able to expend considerable time and effort in making their voice heard as well.

\end{multicols}
\begin{table}[H]
\caption{Diversity metrics for various hashtags}
\label{table2}
\centering
\begin{tabular}{llcccccc}
\toprule
\multicolumn{2}{c}{Hashtag} &
\multicolumn{3}{c}{Full sample} &
\multicolumn{3}{c}{Top decile}\\
\cmidrule(r){1-2}
\cmidrule(r){3-5}
\cmidrule(r){6-8}
Name & Subject type & $H'$ & $\lambda \times 10^4$ & $E_{var}$ & $H'$ & $\lambda \times 10^4$ & $E_{var}$ \\
\midrule

\texttt{\#auspol} & Politics & 8.4148 & 8.6722 & 0.3919 & 5.6269 & 8.6143 & 0.6143 \\
\texttt{\#blacklivesmatter} & Politics & 10.6933 & 1.2001 & 0.7085 & 4.6908 & 1.1666 & 0.7069 \\
\texttt{\#cashinin} & Politics & 6.1643 & 63.5415 & 0.6245 & 2.8426 & 61.3470 & 0.6671 \\
\texttt{\#dataviz} & Professional & 7.6130 & 16.4942 & 0.8492 & 2.3283 & 15.0661 & 0.7112 \\
\texttt{\#ferguson} & Politics & 10.5688 & 2.2599 & 0.6730 & 5.2954 & 2.2405 & 0.6360 \\
\texttt{\#gamergate} & Entertainment & 8.6397 & 6.6783 & 0.3204 & 7.6901 & 6.6778 & 0.2919 \\
\texttt{\#mtvstars} & Entertainment & 6.9324 & 145.6970 & 0.7445 & 3.9151 & 145.6642 & 0.5407 \\
\texttt{\#p2} & Politics & 8.6366 & 17.3569 & 0.5978 & 4.5832 & 17.2662 & 0.6249 \\
\texttt{\#rstats} & Professional & 6.0897 & 47.2607 & 0.8218 & 1.7600 & 39.3447 & 0.7980 \\
\texttt{\#startup} & Professional & 6.9181 & 76.6908 & 0.5778 & 3.8879 & 76.4920 & 0.4257 \\
\texttt{\#tcot} & Politics & 9.3136 & 8.8127 & 0.4223 & 6.2346 & 8.7970 & 0.5940 \\
\texttt{\#uniteblue} & Politics & 8.4368 & 9.8394 & 0.5576 & 4.4738 & 9.6779 & 0.6784 \\

\bottomrule
\end{tabular}
\end{table}
\begin{multicols}{2}

This study focused on two fundamental questions. First, are metrics of population diversity and evenness as they are used in population ecology useful metrics of the diversity of a TDN? Second, what do such metrics say about a particular TDN?

\subsection{Validity of the metrics} 
\label{sub:validity_of_the_metrics}

As discussed above, each of the metrics had their own suitability spectrum. In other words, the metrics considered each had a particular information value. As such, the crucial issue for researchers will be to select the appropriate metric for the research question.

\begin{enumerate}
	\item For \textbf{research questions where richness is relevant}, Shannon's $H'$ metric is a good way to differentiate rich and vibrant TDNs from TDNs that are either dominated by a few loud voices \emph{or} are relatively small.
	\item For \textbf{near-complete independence from $R$ and a good way to highlight small imbalances}, the $\lambda$ metric is a suitable indicator that the conversation is dominated by a few prominent participants.
	\item For the \textbf{assessment of evenness sensitive to changes in the share of all contributors regardless of size}, $E_{var}$ is the best metric. As the expansion showed, it is capable of indicating small changes in the proportion of the least-contributing members as well, making it suitable for sensitive assays of changes in a population's participation rate.
\end{enumerate}


\subsection{Interpretation} 
\label{sub:interpretation}

Just as each of the metrics had a particular aspect of validity, their interpretation in the context of TDNs is subtly different for each.

\begin{enumerate}
	\item \textbf{Interpreting $H'$:} A higher $H'$ indicates a more vibrant conversation, either by increasing diversity (fewer dominant participants and lower dominance rates), increasing the number of participants or both. For the effectiveness assessment of a TND, such as when evaluating the efficiency of social media interventions to generate discourse around a particular topic, this metric is more valuable than the $R$-independent metrics, since richness in and of itself acts as a proxy of reach and constitutes an optimisation goal.
	\item \textbf{Interpreting $\lambda$:} $\lambda$ is an extremely sensitive measure of diversity that is largely independent of $R$. It is a good metric to measure evenness and dominance, and very sensitive to even small increases in the dominance of the largest few contributors. As such, where the research question seeks to identify the relative dominance of the most prominent voices, the $\lambda$ metric is most appropriate.
	\item \textbf{Interpreting $E_{var}$:} Unlike $\lambda$, $E_{var}$ excels at both ends of the user/frequency distribution. It is more sensitive to phenomena such as the decreasing prominence of already less prominent users, a 'silencing' phenomenon that can indicate a TDN's turn from discourse towards information distribution with the occasional comment from other contributors. 
\end{enumerate}




\section{Conclusion} 
\label{sec:conclusion}

Metrics of diversity in population ecology have been usefully applied in other fields, such as competition economics. However, to date, they have not been used in the context of analysing TDNs. This is not the least due to the apparent differences, such as the vastly larger richness and, usually, individuals within TDN samples. This study concluded that ecological metrics of diversity are similarly useful in describing various features of TDNs, but need to be applied within their domains. The conclusion is limited on the evidence from a relatively small number of hashtags, but the sample can for many reasons be regarded as representative. Further research and validation of ecological metrics of social media interactions on TDNs is certainly required and justified, in particular with a view to classification and clustering of TDNs and cross-correlation of value ranges to particular patterns of centre-periphery distributions (such as centrality measures from SNA).




\bibliographystyle{plain}
\bibliography{sdi-article}

\begin{thebibliography}{10}

\bibitem{alatalo1981problems}
Rauno~V Alatalo.
\newblock Problems in the measurement of evenness in ecology.
\newblock {\em Oikos}, pages 199--204, 1981.

\bibitem{bailey1971optimal}
Duncan Bailey and Stanley~E Boyle.
\newblock The optimal measure of concentration.
\newblock {\em Journal of the American Statistical Association},
  66(336):702--706, 1971.

\bibitem{bruns2011new}
Axel Bruns and Jean~E Burgess.
\newblock New methodologies for researching news discussion on twitter.
\newblock 2011.

\bibitem{calkins1983new}
Stephen Calkins.
\newblock The new merger guidelines and the herfindahl-hirschman index.
\newblock {\em California Law Review}, pages 402--429, 1983.

\bibitem{cohen1983herfindahl}
Neil~B Cohen and Charles~A Sullivan.
\newblock Herfindahl-hirschman index and the new antitrust merger guidelines:
  Concentrating on concentration.
\newblock {\em Tex. L. Rev.}, 62:453, 1983.

\bibitem{gonzalez2010structure}
Sandra Gonzalez-Bailon, Andreas Kaltenbrunner, and Rafael~E Banchs.
\newblock The structure of political discussion networks: a model for the
  analysis of online deliberation.
\newblock {\em Journal of Information Technology}, 25(2):230--243, 2010.

\bibitem{hannan1997market}
Timothy~H Hannan.
\newblock Market share inequality, the number of competitors, and the hhi: An
  examination of bank pricing.
\newblock {\em Review of Industrial Organization}, 12(1):23--35, 1997.

\bibitem{highfield2011challenges}
Tim Highfield, Lars Kirchhoff, and Thomas Nicolai.
\newblock Challenges of tracking topical discussion networks online.
\newblock {\em Social Science Computer Review}, 29(3):340--353, 2011.

\bibitem{hunter1988numerical}
Paul~R Hunter and Michael~A Gaston.
\newblock Numerical index of the discriminatory ability of typing systems: an
  application of simpson's index of diversity.
\newblock {\em Journal of clinical microbiology}, 26(11):2465--2466, 1988.

\bibitem{magurran2004measuring}
Anne~E Magurran.
\newblock Measuring biological diversity.
\newblock 2004.

\bibitem{magurran1988ecological}
Anne~E Magurran and Anne~E Magurran.
\newblock {\em Ecological diversity and its measurement}, volume 168.
\newblock Springer, 1988.

\bibitem{marfels1975bird}
Christian Marfels.
\newblock Bird's eye view to measures of concentration, a.
\newblock {\em Antitrust Bull.}, 20:485, 1975.

\bibitem{mason2003index}
Norman~WH Mason, Kit MacGillivray, John~B Steel, and J~Bastow Wilson.
\newblock An index of functional diversity.
\newblock {\em Journal of Vegetation Science}, 14(4):571--578, 2003.

\bibitem{mcintosh1967index}
Robert~P McIntosh.
\newblock An index of diversity and the relation of certain concepts to
  diversity.
\newblock {\em Ecology}, pages 392--404, 1967.

\bibitem{miller1982herfindahl}
Richard~A Miller.
\newblock Herfindahl-hirschman index as a market structure variable: An
  exposition for antitrust practitioners, the.
\newblock {\em Antitrust Bull.}, 27:593, 1982.

\bibitem{pielou1966shannon}
EC~Pielou.
\newblock Shannon's formula as a measure of specific diversity: its use and
  misuse.
\newblock {\em American Naturalist}, pages 463--465, 1966.

\bibitem{pielou1966measurement}
ECJ Pielou.
\newblock The measurement of diversity in different types of biological
  collections.
\newblock {\em Journal of theoretical biology}, 13:131--144, 1966.

\bibitem{pielou1977mathematical}
Evelyn~Chris Pielou.
\newblock {\em Mathematical ecology}.
\newblock John wiley et sons, 1977.

\bibitem{porter2004typology}
Constance~Elise Porter.
\newblock A typology of virtual communities: A multi-disciplinary foundation
  for future research.
\newblock {\em Journal of Computer-Mediated Communication}, 10(1):00--00, 2004.

\bibitem{rhoades1993herfindahl}
Stephen~A Rhoades.
\newblock Herfindahl-hirschman index, the.
\newblock {\em Fed. Res. Bull.}, 79:188, 1993.

\bibitem{shannon1993collected}
Claude Shannon.
\newblock Collected papers.
\newblock 1993.

\bibitem{smith1996consumer}
Benjamin Smith and J~Bastow Wilson.
\newblock A consumer's guide to evenness indices.
\newblock {\em Oikos}, pages 70--82, 1996.

\bibitem{weinstock1982using}
David~S Weinstock.
\newblock Using the herfindahl index to measure concentration.
\newblock {\em Antitrust Bull.}, 27:285, 1982.

\end{thebibliography}



\end{multicols}

\end{document}